\documentclass[a4paper,12pt]{article}

\usepackage{amsmath, amssymb, authblk, a4wide, bm, cancel, color, graphicx, subfig, enumerate, appendix, youngtab, cite, hyperref, bbold, verbatim, slashed}

\newcommand{\pdiff}[2]{\frac{\partial#1}{\partial#2}}
\newcommand{\dimint}[2]{\int\mathrm{d}^{#1}#2\,}
\newcommand{\dimintlim}[4]{\int_{#3}^{#4}\mathrm{d}^{#1}#2\,}
\newcommand{\nbrack}[1]{\left(#1\right)}
\newcommand{\sbrack}[1]{\left[#1\right]}
\newcommand{\expect}[1]{\langle#1\rangle}
\newcommand{\norm}[1]{\left|#1\right|}
\newcommand{\pfrac}[2]{\left(\frac{#1}{#2}\right)}

\def\be{\begin{equation}} \def\ee{\end{equation}}
\def\ba{\begin{eqnarray}} \def\ea{\end{eqnarray}} \def\cJ{{\cal J}}
\def\cL{{\cal L}}

\def\cO{{\cal O}}

\def\uno{\mbox{1 \kern-.59em {\rm l}}}

\numberwithin{equation}{section}
\numberwithin{figure}{section}
\numberwithin{table}{section}

\begin{document}
\title{\vspace{2cm}
\Large{\textbf{Radiative corrections to the composite Higgs mass from a gluon partner}}}
\author[1]{\small{\bf James Barnard}\thanks{\texttt{james.barnard@unimelb.edu.au}}}
\author[1,2]{\small{\bf Tony Gherghetta}\thanks{\texttt{tgher@unimelb.edu.au}}}
\author[1]{\small{\bf Anibal Medina}\thanks{\texttt{anibal.medina@unimelb.edu.au}}}
\author[1]{\small{\bf Tirtha Sankar Ray}\thanks{\texttt{tirtha.sankar@unimelb.edu.au}}}
\affil[1]{ARC Centre of Excellence for Particle Physics at the Terascale, School of Physics, The University of Melbourne, Victoria 3010, Australia }
\affil[2]{Stanford Institute of Theoretical Physics, Stanford University, Stanford, CA 94305, USA}
\date{}
\maketitle

\begin{abstract}
\baselineskip=15pt
\noindent 
In composite Higgs models light fermionic top partners often play an
important role in obtaining a 126 GeV Higgs mass.  The presence of
these top partners implies that coloured vector mesons, or massive
gluon partners, most likely exist. Since the coupling between the top
partners and gluon partners can be large there are then sizeable two-loop contributions to the composite Higgs mass.  We compute the
radiative correction to the Higgs mass from a gluon partner in the
minimal composite Higgs model and show that the Higgs mass is in fact
reduced.  This allows the top partner masses to be increased, easing
the tension between having a light composite Higgs and heavy top
partners.
\end{abstract}

\newpage
\section{Introduction}

The recent discovery of the Higgs boson at the Large Hadron Collider
(LHC) \cite{Chatrchyan:2012ufa, Chatrchyan:2013lba, Aad:2012tfa}
confirms that the Higgs mechanism is responsible for spontaneously
breaking electroweak symmetry in the Standard Model (SM)\@. However,
the question of whether the Higgs sector is natural or not remains to
be answered. The two most appealing solutions for stabilising the weak
scale are supersymmetry (see ref.~\cite{Martin:1997ns} for a review)
and compositeness \cite{Kaplan:1983fs, Georgi:1984af}. Both are now
constrained by measurements of the Higgs mass and its couplings, which
have led to consequences for the spectrum of exotic states predicted
in these two scenarios. In supersymmetric models the radiative
corrections from coloured states, such as the stop and gluino, have
been extensively studied in the literature and shown to have an
important effect on tuning in the Higgs sector.  Combined with the
lower limits on sparticle masses from the LHC, the conclusion is that
supersymmetric models are now tuned to below the 5$\%$ level
\cite{Gherghetta:2012gb}.

Exotic coloured states also exist in composite Higgs models and can
similarly play an important role in determining the Higgs mass.  In
many of these models the SM Higgs doublet is identified with
Nambu-Goldstone modes, arising from the spontaneous breaking of a
global symmetry group $G$ to a subgroup $H$ due to some strong
dynamics.  However, the original global symmetry is also explicitly
broken via mixing between operators in the strongly coupled sector and
elementary fields in the SM sector, as the latter need not come in
complete representations of $G$. Hence a Coleman-Weinberg type
potential is generated, leading to dynamical electroweak symmetry
breaking and a mass for the Higgs boson \cite{Contino:2003ve}.

A key component in such models is the existence of composite,
fermionic top partner resonances in the low energy spectrum.  They are
required to facilitate a strong coupling between the top quark and the
composite Higgs (through substantial mixing between composite and
elementary states in the top sector) and, often, to break electroweak
symmetry in the first place.  The scale of the Higgs mass is typically
set by these top partner masses so, to provide a Higgs mass of around
126 GeV \cite{Matsedonskyi:2012ym,
  Marzocca:2012zn,Pomarol:2012qf,Panico:2012uw,Pappadopulo:2013vca},
the top partners cannot be too heavy.  Including such coloured
fermionic resonances generically implies that there will also be
coloured vector meson resonances in the low energy spectrum. This
follows because, even though a complete description of the underlying
dynamics of the strong sector remains unknown, the constituents of the
strong sector must be coloured in order to produce top partner bound
states charged under $SU(3)$ colour. Consequently, the strong sector
is also expected to produce coloured vector mesons that necessarily
couple to any top partner bound state.  We will refer to these states
as gluon partners.  Indeed, if the strong sector contains a fermionic
operator $\cO_\psi$, in the fundamental representation of $SU(3)$ so
as to produce top partners, one can always write down a vector
operator $\bar{\cO_\psi}\gamma^\mu{\cO}_\psi$ in the adjoint
representation of $SU(3)$.  Equivalently, in the five-dimensional (5D)
version of these composite Higgs models
\cite{Agashe:2004rs,Medina:2007hz,Gripaios:2009pe, Mrazek:2011iu,
  Bertuzzo:2012ya} the gluon partners are simply the Kaluza-Klein
gluons, which are required by 5D gauge invariance if the SM fermions
are located in the bulk.

Thus far the contribution of gluon partners to the Higgs mass has been
neglected. Since they only couple to the Higgs through the top
partners, the size of the correction to the Higgs mass will be
proportional to the gluon partner-top partner coupling, $\alpha_G$.
This coupling can be estimated either through direct calculations in
the 5D theory \cite{Carena:2007tn} or via holographic techniques
\cite{Gherghetta:2011na}. In both instances the coupling is large,
hence the contribution to the Higgs mass is expected to be sizeable.

In this paper we explicitly compute the leading order correction and
do indeed find it to be important.  Specifically, we calculate the
one-loop correction to the two-point function of (Dirac) top partners
coming from a massive gluon partner. This first result is model
independent but, to quantify the effect on the Higgs mass, we apply it
to the specific example of the MCHM$_{\bf 5}$\@.  Interestingly, we
find that the Higgs mass is decreased in this particular model,
thereby easing some of the tension between the tuning in composite
Higgs models and the non-observation of coloured top partners.  For a
gluon partner of mass 3 TeV, a spontaneous global symmetry breaking
scale of about 750 GeV and with the Higgs mass fixed at 126 GeV, the
top partners are about 10\% heavier than in the uncorrected model.

The rest of this paper is organised as follows. In section 2 we
consider gluon partners in a general composite Higgs model. The
one-loop radiative correction from a gluon partner to the two-point
function is first estimated in the large $N$ limit and then the exact
calculation is performed with the final result expressed in integral
form.  The exact result is used in section 3 to compute the
contribution to the composite Higgs mass in the MCHM$_{\bf 5}$\@.  The
concluding remarks are presented in section 4. In appendix A we use
the holographic basis to estimate the size of the radiative
correction, and in appendix B we list the Passarino-Veltman integral
expressions utilised earlier in the paper.

\section{Gluon partners and the composite Higgs mass}

Many composite Higgs models consists of a strong sector that is
responsible for producing a set of Nambu-Goldstone bosons, a
combination of which is identified with the Higgs boson, $h$.  The
strong sector is joined by an elementary sector containing the SM
fermions and gauge bosons, and the two sectors are assumed to mix only
through bilinear couplings \be\label{eq:Lmix} \cL_{\rm
  mix}=\bar{q}_L\cO_R+\bar{t}_R\cO_L+A_g^\mu\cJ_\mu+\ldots \ee where
the elementary fields $q_L, t_R$ and $A_g$ are the left handed top
quark doublet, right handed top and gluon respectively.  We have only
included the fermionic operators $\cO_{L/R}$ and the strong sector
$SU(3)$ colour current $\cJ$; the dots represent other bilinear
couplings which will not be considered here, such as those for the
light fermions.

Integrating out the strong sector results in an effective Lagrangian
for the top-Higgs sector \cite{Agashe:2004rs}
\begin{align}\label{eq:Leff}
\cL_{\rm eff}={} & \bar{t}_L\slashed{p}\sbrack{\Pi_L^0(p^2)+{\cal
    Y}_L(h/f)\Pi_L^h(p^2)}t_L+\bar{t}_R\slashed{p}\sbrack{\Pi_R^0(p^2)+{\cal
    Y}_R(h/f)\Pi_R^h(p^2)}t_R+{} \nonumber\\ & \sbrack{\bar{t}_L{\cal
    Y}_M(h/f)M(p^2)t_R+\mbox{h.c.}},
\end{align}
where $f$ is the Nambu-Goldstone boson decay constant.  The functions
$\Pi$ and $M$ are determined by two-point functions of the fermionic
operators
\begin{align}
\slashed{p}\Pi_{L/R}(p^2) & \sim\expect{\cO_{L/R}(p)\bar{\cO}_{L/R}(-p)}, &
M(p^2) & \sim\expect{\cO_L(p)\bar{\cO}_R(-p)},
\end{align}
up to a factor of $+1$ in $\Pi^0$ coming from the usual SM kinetic
term (i.e.\ $\Pi^0\sim1+\expect{\cO\bar{\cO}}$).  Assuming the strong
sector is a large $N$ gauge theory and working to leading order in
$1/N$, one can write these two-point functions as a sum over narrow,
top partner resonances, $Q_n$, to find
\begin{align}\label{eq:Mdef}
\Pi_{L/R}(p^2) & =\sum_{n=1}^\infty\frac{a_n|F^{L/R}_n|^2}{p^2-m_{Q_n}^2}, &
M(p^2) & =\sum_{n=1}^\infty\frac{b_nF^L_nF^R_n{}^*m_{Q_n}}{p^2-m_{Q_n}^2},
\end{align}
for masses $m_{Q_n}$, constant form factors $F_n$ and where the
coefficients $a_n,b_n$ are derived from the group structure of any
particular model.  Their precise form and that of their prefactors,
the functions ${\cal Y}(h/f)$ of the Higgs fields, are determined by
the details of the global symmetry breaking pattern and the
representations into which the top quarks are embedded.  However, they
can always be split into those components that are sensitive to the
Higgs vacuum expectation value (VEV) and those that are not, hence
${\cal Y}_L(0)={\cal Y}_R(0)={\cal Y}_M(0)=0$.

Since the top quarks do not make up full representations of the global
symmetry group in the strong sector, the symmetry is explicitly broken
and a potential can be generated for the erstwhile Nambu-Goldstone
bosons.  At one loop the potential from the top sector is given by
\begin{align}
V_{\rm
  eff}(h)=-2N_c\int\frac{\mathrm{d}^4p_E}{(2\pi)^4}\,\ln\left(-p_E^2\sbrack{\Pi^0_L(p_E^2)+{\cal
    Y}_L(h/f)\Pi_L^h(p_E^2)}\sbrack{\Pi^0_R(p_E^2)+{\cal
    Y}_R(h/f)\Pi_R^h(p_E^2)}\right. \nonumber\\ \left.{}-\norm{{\cal
    Y}_M(h/f)M(p_E^2)}^2\right),
\end{align}
where $N_c=3$ is the QCD colour factor and the integral is performed
over Euclidean momentum $p_E$.  Expanding the logarithm and discarding
the constant term gives an approximate form for the potential
\begin{align}\label{eq:Vh}
V_{\rm
  eff}(h)\approx-6\int\frac{\mathrm{d}^4p_E}{(2\pi)^4}\,\left[\frac{{\cal
      Y}_L(h/f)\Pi_L^h(p_E^2)}{\Pi^0_L(p_E^2)}+\frac{{\cal
      Y}_R(h/f)\Pi_R^h(p_E^2)}{\Pi^0_R(p_E^2)}+\frac{\norm{{\cal
        Y}_M(h/f)M(p_E^2)}^2}{p_E^2\Pi_L^0(p_E^2)\Pi_R^0(p_E^2)}\right. \nonumber\\ \left.{}-\pfrac{{\cal
      Y}_L(h/f)\Pi_L^h(p_E^2)}{2\Pi^0_L(p_E^2)}^2-\pfrac{{\cal
      Y}_R(h/f)\Pi_R^h(p_E^2)}{2\Pi^0_R(p_E^2)}^2\right],
\end{align}
which can be considered an expansion in ${\cal Y}(h/f)$ (corresponding
to a small Higgs VEV) or an expansion in $\Pi^h_{L/R}$ and $M$
(corresponding to weak mixing between elementary and composite degrees
of freedom).

\subsection{Gluon partner contributions\label{sec:NDA}}

The operators $\cO_L$ and $\cO_R$ in the strong sector, which are
required to mix with the top quark, guarantee the existence of top
partners in this framework.  However, as argued in the introduction,
they are typically accompanied by massive, coloured vector meson
resonances, or gluon partners.  These are associated with the strong
sector current operator $\cJ$ as, at leading order in $1/N$, the
two-point function $\langle \cJ\cJ \rangle$ can be written as a sum
over narrow gluon resonances, $G_n$, much like the two-point functions
$\Pi$ and $M$.

Because the Higgs is colour neutral, any correction to its mass from
gluon partners must enter,
at two-loop order, through the two-point functions of $\cO_L$
and $\cO_R$.\footnote{There is also a vertex correction but this is
  heavily suppressed compared to the contribution from the two-point
  function, as the Higgs is a Nambu-Goldstone mode so can only couple
  through derivatives in the strong sector.  Hence the
  $\expect{h\cO_L\cO_R}$ three-point function (which already comes
  with an extra $1/\sqrt{N}$ suppression) must vanish and the vertex
  correction is suppressed by several additional composite-elementary
  mixing factors.}  This means the effect can be accounted for by
calculating the gluon partner corrections to the functions $M$ and
$\Pi$, defined in eq.~\eqref{eq:Mdef} in the large $N$ limit.  A naive
large $N$ analysis suggests that the correction is not important
because it depends on a $\cO\cO\cJ$ coupling in the strong sector,
which scales like $1/\sqrt{N}$.  However, this is only a scaling
dependence and ignores the prefactor for the coupling, which we will
now show is large.  To estimate the importance of the correction more
accurately we will use the AdS/CFT correspondence to estimate the
strength of couplings in the strong sector, then use large $N$ results
to estimate the amplitudes of the relevant diagrams.

In theories of warped extra dimensions one can relate the $N$
appearing in the large $N$ CFT expansion with the 5D gauge couplings
using \cite{Gherghetta:2011na} \be\label{eq:Ngauge}
\frac{1}{\kappa_iN}=\frac{g_{5,i}^2k}{16\pi^2}=\frac{g_{i}^2}{16\pi^2}\ln\pfrac{\Lambda_{\rm
    UV}}{\Lambda_{\rm IR}}, \ee where $k$ is the curvature scale of
the 5D warped AdS space, $g_{5,i}$ a bulk gauge coupling, $g_i$ a
four-dimensional (4D) gauge coupling and $\kappa_i$ a numerical factor
distinguishing between different gauge groups.  We have also made use
of the relation between 5D and 4D gauge couplings
$g_{5,i}^2k=g_{i}^2\ln{(\Lambda_{\rm UV}/\Lambda_{\rm IR})}$, which
includes the logarithmic running between the UV and the IR scales.
The expression \eqref{eq:Ngauge} can be used to provide quantitative
information about the strength of the couplings between resonances in
the strong sector.

To estimate the coupling between the top partner and the gluon
partner, we first use the $SU(3)$ gauge coupling to obtain the value
for $\kappa_3 N$.  Using \eqref{eq:Ngauge} one finds \be\label{eq:gN}
\frac{1}{\kappa_3 N} =\frac{g_3^2}{16\pi^2}\ln\pfrac{\Lambda_{\rm
    UV}}{\Lambda_{\rm IR}}\approx\frac{3}{4\pi}, \ee where
$\alpha_3\approx0.1$ is the QCD gauge coupling strength and
$\ln\left(\Lambda_{\rm UV}/\Lambda_{\rm IR}\right)\approx 30$.  Thus
the coupling between the fermionic and gluon resonances can be
estimated as \be\label{eq:alphaG}
\alpha_G\simeq\frac{4\pi}{\kappa_3N}\approx3.  \ee This compares well
with the exact 5D calculation \cite{Carena:2007tn}, where the overlap
integral between the first Kaluza-Klein gluon and the first
Kaluza-Klein top gives $\alpha_G\simeq 2.1$.\footnote{Due to the
  subtleties of fermion boundary conditions and localisation in the
  extra dimension, we focus on figure 2 of ref.~\cite{Carena:2007tn}.
  The case most relevant to the discussion here corresponds to a large
  positive value of the bulk mass parameter $c_1$, which implies
  $g_{G^1t^1_Lt^1_L}=g_{G^1t^1_Rt^1_R}\approx-5 g_{s}(f)$ where $
  g_{s}(f)\simeq 1.02$ is the strong coupling evaluated at the scale
  $f$.  In this region of parameter space one recovers the usual
  Kaluza-Klein fermion spectrum for unmixed boundary conditions.}
Note that in the estimate \eqref{eq:gN} we have neglected SM loop and
brane kinetic term contributions to the 4D coupling.  These
contributions are model dependent and can increase or decrease the
coupling $\alpha_G$ \cite{Csaki:2008zd, Agashe:2008uz}.  For example,
if the SM loop contributions are not cancelled by UV brane kinetic
terms then the value \eqref{eq:alphaG} of the coupling $\alpha_G$ is
reduced by approximately 1/4.\footnote{We thank Kaustubh Agashe for
  bringing this point to our attention.} To avoid model dependency we
will assume the value \eqref{eq:alphaG} for concreteness in the rest
of this paper.

\begin{figure}[!t]
\begin{center}
\includegraphics[width=0.4\textwidth]{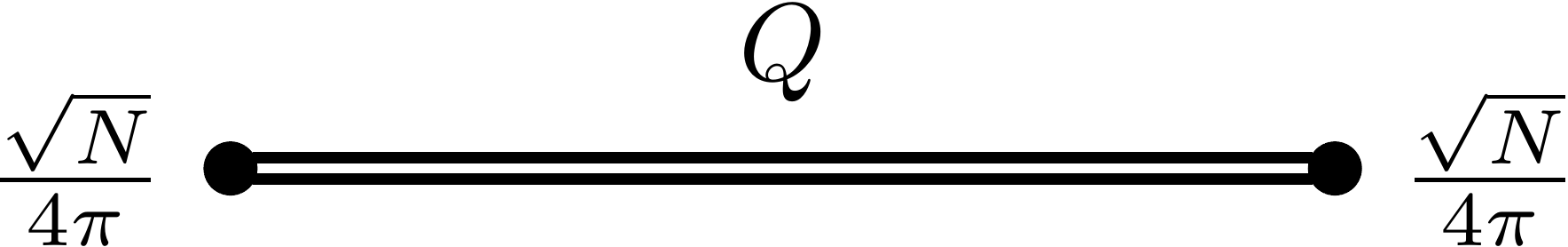}\\\vspace{5mm}
\includegraphics[width=0.4\textwidth]{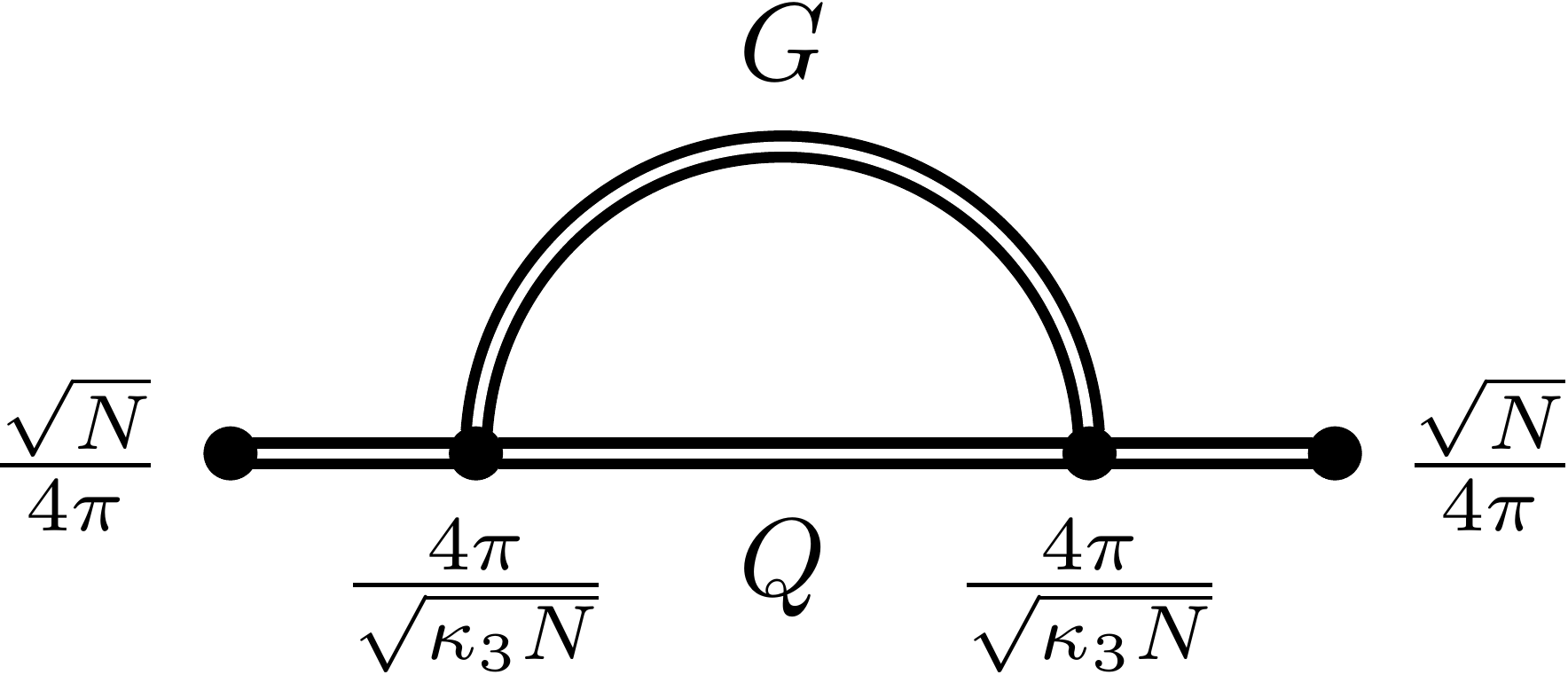}
\caption{The tree-level (top) and one-loop contribution (bottom) to
  the two-point function of the fermionic operators in the large $N$
  limit.\label{fig:selfenergy}}
\end{center}
\end{figure}

Now the coupling \eqref{eq:alphaG} can be used to estimate the size of
the one-loop gluon partner contribution to the two-point functions
relative to that of the tree-level contribution: \be\label{eq:ratio}
\frac{\mbox{one-loop}}{\mbox{tree}}=\frac{(N/16\pi^2)\times(16\pi^2/\kappa_3N)\times(C_2(N_c)/16\pi^2)}{(N/16\pi^2)}\approx\frac{1}{\pi}.
\ee The expression on the left hand side is obtained using the large
$N$ result $\expect{\mathcal{O}_{L}\mathcal{O}_{R}}\sim N/16\pi^2$ for
the tree-level contribution in the denominator (the top diagram in
figure \ref{fig:selfenergy}).  For the one-loop gluon partner
contribution in the numerator (the bottom diagram in figure
\ref{fig:selfenergy}) there are two vertices coupling top partners to
a gluon partner, each contributing a factor of $4\pi/\sqrt{\kappa_3
  N}$, an additional loop factor $1/16\pi^2$, as well as the quadratic
Casimir $C_2(3)=4/3$ (coming from $t^at^a=C_2(N_c)\uno$, where $t^a$
are the generators of $SU(3)$ in the fundamental representation).
Finally, we substitute in the estimated value for $\kappa_3$ from
eq.~\eqref{eq:gN} to get a numerical value.

Despite being a higher order effect, we find that the contribution due
to the gluon partners could be of order $30\%$.  An alternative
derivation, based on the mixing between the holographic and mass bases
and yielding the same result, is presented in the appendix.  Note that
these estimates neglect a momentum dependent loop function, which will
be explicitly computed in the next section.

\subsection{Exact calculation}

\begin{figure}[!t]
\begin{center}
\includegraphics[width=0.27\textwidth]{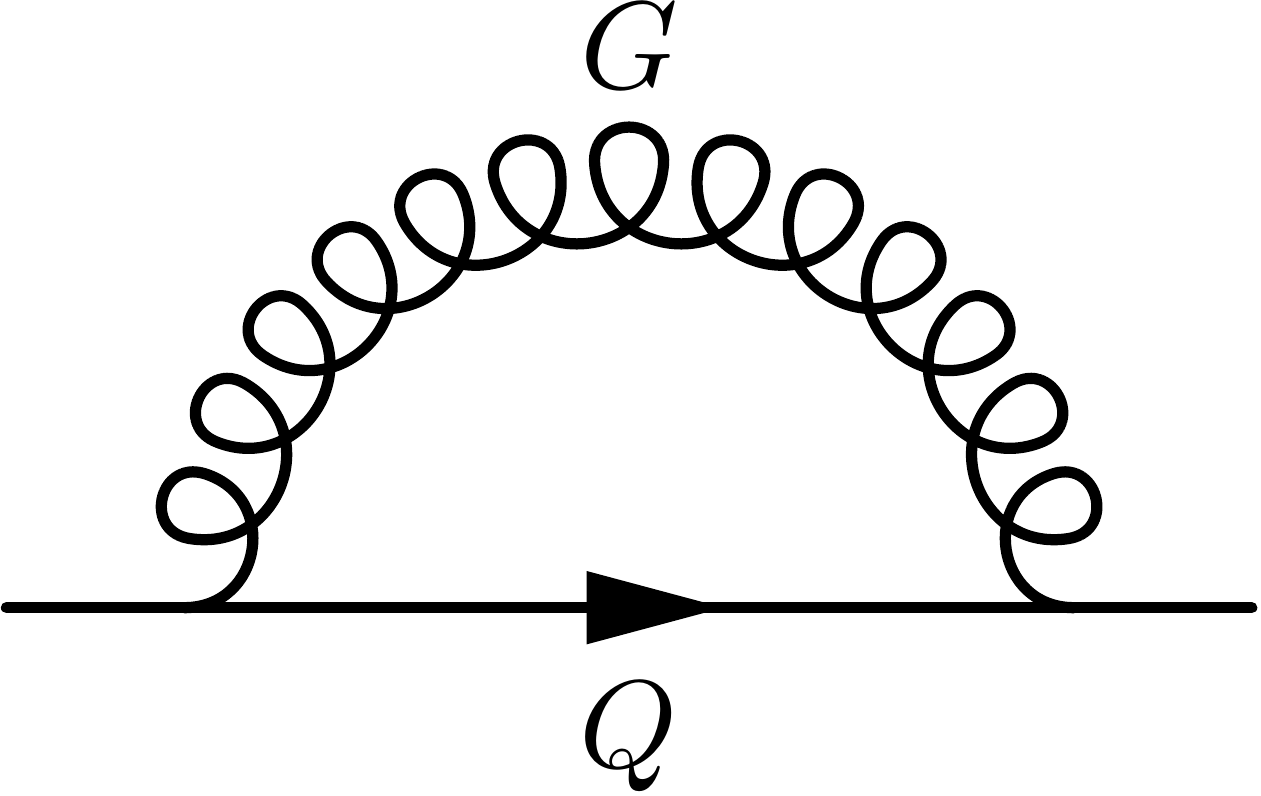}
\caption{The one-loop contribution to the self energy of a fermion
  resonance $Q$ from a gluon partner $G$.\label{fig:selfenergyC}}
\end{center}
\end{figure}

To fully quantify the effect of gluon partners on the Higgs mass we
must calculate the contribution from figure \ref{fig:selfenergyC}
explicitly.  This assumes that the gluon partners associated with the
current $\cJ$ can be modelled as narrow resonances, like their
fermionic brethren.  Each top partner propagator is renormalised to
\be\label{eq:Sr} S^{-1}(p)=S^{-1}_0(p)+i\Sigma_1(p), \ee at one-loop
order, where $S^{-1}_0(p)=-i(\slashed{p}-m_Q)$ is the unrenormalised
propagator and $\Sigma_1(p)$ is the one-loop renormalised self energy.
Including appropriate counterterms for the wave function and mass
renormalisation, this self energy is given by \be\label{eq:Sigmar}
\Sigma_1(p)=\Sigma(p)-(\slashed{p}-m_Q)\delta Z_2-\delta m_Q, \ee
where \be\label{eq:Sigmal}
i\Sigma(p)=\frac{16\pi}{3}\alpha_G\int\frac{{\rm
    d}^4k}{(2\pi)^4}\,\gamma^\mu\frac{1}{\slashed{k}-m_Q}\gamma^\nu\frac{\eta_{\mu\nu}}{(k-p)^2-m_G^2},
\ee is the integral expression obtained from the diagram in figure
\ref{fig:selfenergyC}.

Renormalising the propagator in the on-shell scheme gives the two
renormalisation conditions
\begin{align}
\left.\Sigma_1(p)\right|_{\slashed{p}=m_Q} & =0, &
\left.\pdiff{\Sigma_1(p)}{\slashed{p}}\right|_{\slashed{p}=m_Q} & =0,
\end{align}
which are used to determine the two counter terms $\delta Z_2$ and
$\delta m_Q$.  By writing \be \Sigma(p)=m_QA(p^2)+\slashed{p}B(p^2),
\ee then solving the above equations and substituting into
eqs.~\eqref{eq:Sr} and \eqref{eq:Sigmar}, we find the one-loop
expression \be\label{eq:prop}
iS^{-1}(p)=\nbrack{1-\widehat{B}(p^2)+\left.m_Q\pdiff{}{\slashed{p}}\sbrack{\widehat{A}(p^2)+\widehat{B}(p^2)}\right|_{\slashed{p}=m_Q}}\nbrack{\slashed{p}-m_Q\sbrack{1+\widehat{A}(p^2)+\widehat{B}(p^2)}},
\ee where we have further defined
\begin{align}
\widehat{A}(p^2) & =A(p^2)-A(m_Q^2), & \widehat{B}(p^2) &
=B(p^2)-B(m_Q^2).
\end{align}
The first factor in \eqref{eq:prop} is absorbed into an overall
rescaling of the top partner fields and is not important for our
purposes.  The second term can be expressed in the propagator as an
effective correction to the top partner mass \be \Delta
m_Q(p^2)=m_Q\sbrack{\widehat{A}(p^2)+\widehat{B}(p^2)}.  \ee We now
need to evaluate eq.~\eqref{eq:Sigmal} to determine the functions
$\widehat{A}(p^2)$ and $\widehat{B}(p^2)$.  The answer is succinctly
expressed in terms of Passarino-Veltman integrals
\cite{Passarino:1978jh} (see the appendix)
\begin{align}
i\Sigma(p) & =\frac{2\alpha_G}{3\pi^3}\int{\rm
  d}^4k\,\frac{2m_Q-\slashed{k}}{[k^2-m_Q^2][(k-p)^2-m_G^2]}
\nonumber\\ &
=\frac{2i\alpha_G}{3\pi}\sbrack{2m_QB_0(p^2,m_Q^2,m_G^2)+\slashed{p}B_1(p^2,m_Q^2,m_G^2)},
\end{align}
and therefore
\begin{align}
\widehat{A}(p^2) &
=\frac{4\alpha_G}{3\pi}\sbrack{B_0(p^2,m_Q^2,m_G^2)-B_0(m_Q^2,m_Q^2,m_G^2)},
\\ \widehat{B}(p^2) &
=\frac{2\alpha_G}{3\pi}\sbrack{B_1(p^2,m_Q^2,m_G^2)-B_1(m_Q^2,m_Q^2,m_G^2)}.
\end{align}
In an explicit integral form we find the final result
\be\label{eq:Deltam} \Delta
m_Q(p^2)=\frac{2\alpha_G}{3\pi}m_Q\dimintlim{}{x}{0}{1}(x-2)\ln\sbrack{\frac{(1-x)m_Q^2+xm_G^2-x(1-x)p^2}{(1-x)^2m_Q^2+xm_G^2}}.
\ee

This one-loop result can be easily generalised to include
contributions from multiple gluon partners by replacing the above
expression \eqref{eq:Deltam} with a sum of identical terms, each using
different values for $\alpha_G$ and $m_G$. However, we expect that the
lightest gluon partner will always dominate because the loop integral
can be seen to decrease as the gluon partner mass is increased, and
the coupling is expected to behave similarly (this latter effect can
be seen explicitly in the 5D calculation).

\subsection{Electroweak and other contributions}

In addition to the coloured contribution we can estimate the
equivalent non-coloured contribution arising from the electroweak
resonances. Using the relation \eqref{eq:Ngauge} we obtain
\be\label{eq:gNW} \frac{1}{N}
=\frac{g_{2}^2}{16\pi^2}\ln\pfrac{\Lambda_{\rm UV}}{\Lambda_{\rm
    IR}}\approx\frac{1}{4\pi}, \ee where $\alpha_{2}\approx1/30$ is
the electroweak coupling strength and, for simplicity, we have set
$\kappa_2=1$, corresponding to a redefinition of $N$.  This enables us
to estimate the coupling $\alpha_\rho$ of the top partner fermions
with the electroweak vector mesons.  Using eq.~\eqref{eq:gNW} we
obtain $\alpha_\rho=4\pi/N=1$.  This is a factor of three smaller than
the gluon coupling \eqref{eq:alphaG}.  The ratio of the electroweak
resonance correction relative to that of the gluon partner correction
is then $[\alpha_\rho C_2(2)]/[\alpha_GC_2(3)] \simeq 3/16$, where we
have further used the $SU(2)$ quadratic Casimir $C_2(2)=3/4$. Thus the
correction coming from electroweak vector mesons is much less
important than the gluon partner correction, and will henceforth be
neglected.\footnote{It should be noted that the electroweak resonances
  are required to restore unitarity.  In principle the gluon and
  electroweak resonance masses are of the same order, hence unitarity
  implies that the gluon partners cannot be arbitrarily heavy.
  However when $v^2=0.1 f^2$, for example, the mass limit on the
  electroweak resonances is not that stringent
  \cite{Bellazzini:2012tv} so there is no real limit on the gluon
  partner masses coming from this observation.}

Having gone beyond leading order in $1/N$ one may anticipate that
other corrections to the Higgs potential, beyond the simple mass shift
of eq.~\eqref{eq:Deltam}, should be considered.  While such
corrections do exist, and are at the same order in $1/N$, they will
always be subdominant.  The reason is that the correction calculated
above is the only one proportional to the top partner-gluon partner
coupling $\alpha_G$, in contrast to other corrections which go like
$\sqrt{\alpha_G\alpha_3}$ or just $\alpha_3$.  The couplings in the
coloured sector satisfy $\alpha_G\gg\alpha_3$, as shown in the
previous subsection, hence any other corrections coming from this
sector can be neglected.  The situation is even more acute in the
non-coloured sector, as these corrections suffer a similar suppression
from the reduced coupling strength, and do not even benefit from QCD
multiplicity factors when evaluating loops.  Another correction one
might consider beyond leading order in $1/N$ is a wave function
renormalisation of the Higgs.  However, in this framework the Higgs
always appears in the ratio $h/f$, meaning that any wave function
renormalisation is simply absorbed in a rescaling of the symmetry
breaking scale.

While the above arguments are robust in the models we are considering
here, where the strong sector only communicates with elementary fields
through bilinear couplings like those in eq.~\eqref{eq:Lmix}, they do
not immediately apply in more general models.  Then, there may be
other corrections to the Higgs mass of similar importance.  However,
it should be noted that the effective Lagrangian \eqref{eq:Leff} will
no longer apply either.  Such models are beyond the scope of this
paper, but could result in interesting deviations from the usual
behaviour found in composite Higgs models.

\section{Effect on the minimal composite Higgs mass}

\begin{figure}[!t]
\begin{center}
\includegraphics[width=0.48\textwidth]{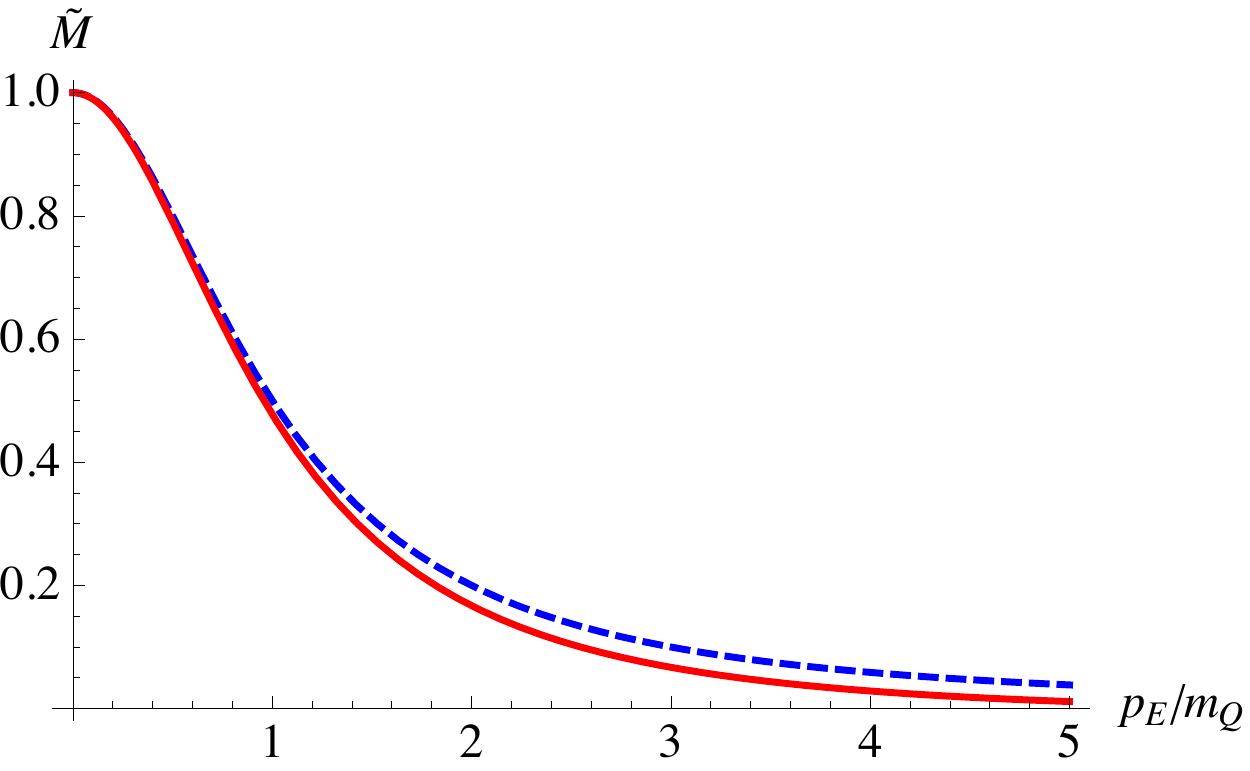}
\includegraphics[width=0.48\textwidth]{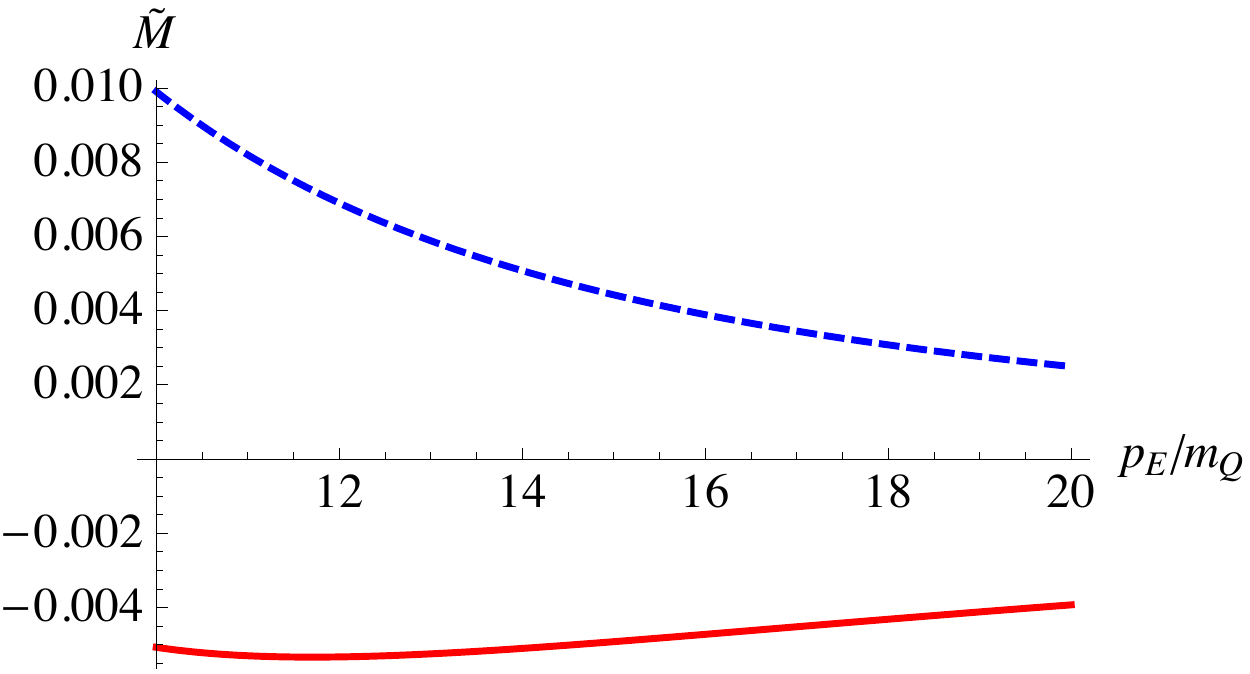}\\\vspace{5mm}
\includegraphics[width=0.48\textwidth]{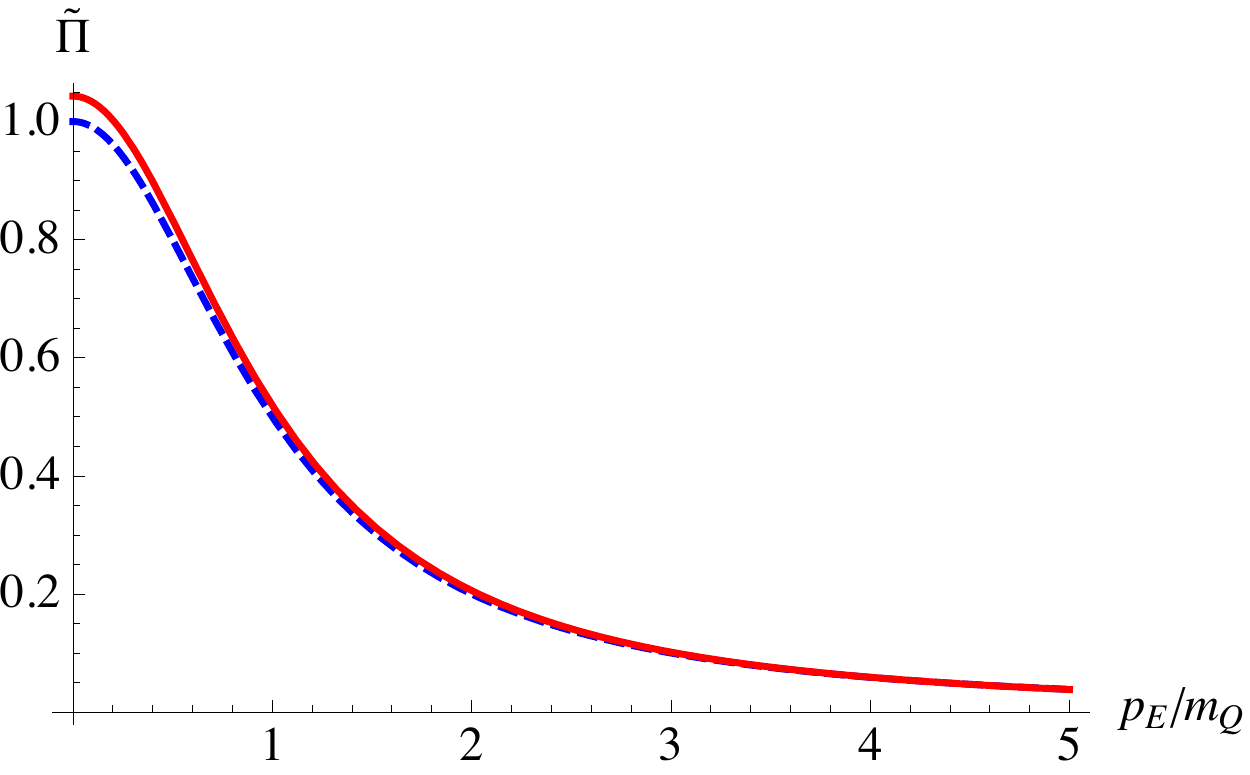}
\includegraphics[width=0.48\textwidth]{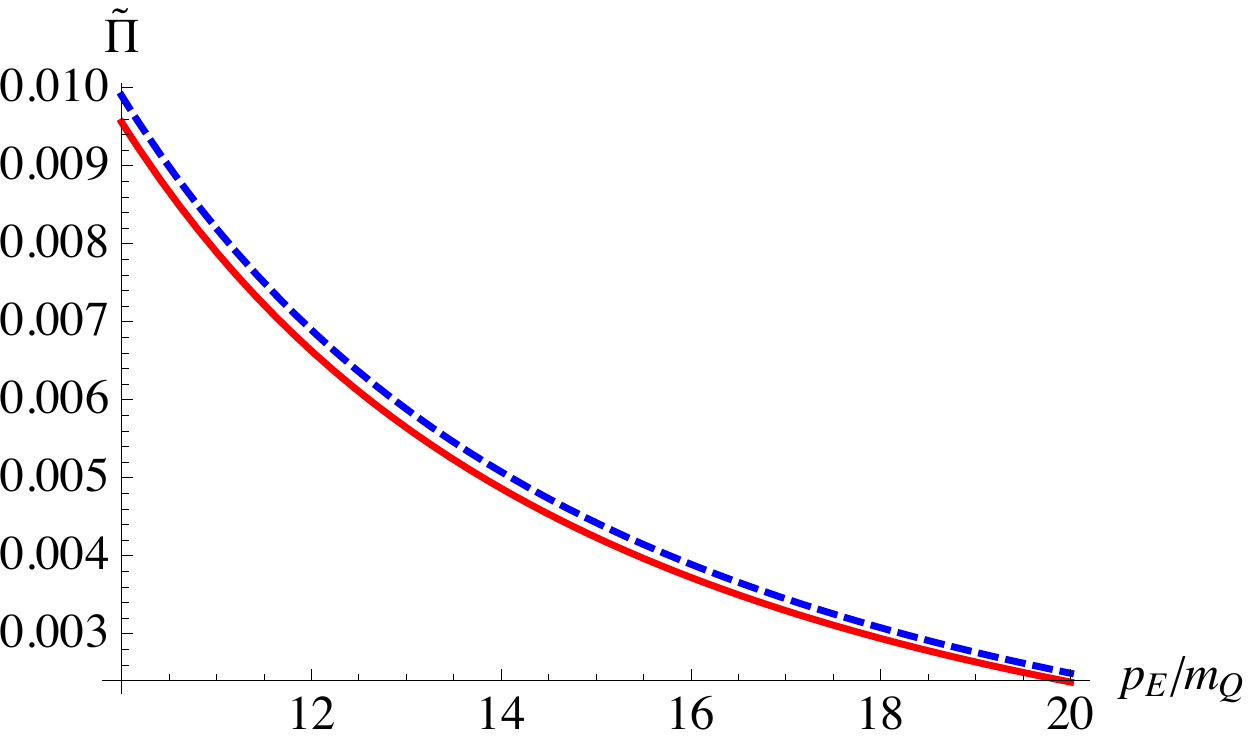}
\caption{The effect of the gluon partner correction on the normalised
  form factors $\widetilde{M}(p_E^2)\equiv v/(\sqrt{2}fm_t)M(p_E^2)$
  (top) and $\widetilde{\Pi}(p_E^2)\equiv(m_Q/M(0))\Pi(p_E^2)$
  (bottom).  The functions are evaluated using eq.~\eqref{eq:Mdefr}
  for a single top and gluon partner resonance, fixing $a=b=1$ and
  $F_L=F_R$ for simplicity.  The normalisation is chosen to respect
  the top mass constraint in eq.~\eqref{eq:mt}.  The dashed lines show
  the uncorrected functions and the solid lines show the effect of
  gluon partner correction for $\alpha_G=3$ and $m_G=3$
  TeV.\label{fig:Mplot}}
\end{center}
\end{figure}

The leading order effect of gluon partners on the Higgs mass in the
large $N$ limit is accounted for by shifting the mass parameters in
eq.~\eqref{eq:Mdef} using the expression in eq.~\eqref{eq:Deltam}.
One then uses the renormalised functions
\begin{align}\label{eq:Mdefr}
M(p^2) & =\sum_{n=1}^\infty\frac{b_nF^L_nF^R_n{}^*(m_{Q_n}+\Delta m_{Q_n}(p^2))}{p^2-(m_{Q_n}+\Delta m_{Q_n}(p^2))^2}, \nonumber\\
\Pi(p^2)_{L/R} & =\sum_{n=1}^\infty\frac{a_n|F^{L/R}_n|^2}{p^2-(m_{Q_n}+\Delta m_{Q_n}(p^2))^2},
\end{align}
to calculate the Higgs potential in eq.~\eqref{eq:Vh}.  Leaving these
functions as they are, i.e.\ not expanding in $\alpha_G$ again,
ensures that the approximation remains good at high momentum, and
corresponds to consistently resumming all contributions generated by
the 1PI diagram in figure \ref{fig:selfenergyC}.  This is easily seen
to be the leading order effect.  Note also that, since the correction
shows up as a mass shift only, all of the group structure is preserved
and we can simply apply existing results to calculate the Higgs mass
with no need to worry about the theory becoming non-renormalisable.
This may not have been the case if the renormalisation procedure had
introduced momentum dependence into the $F$'s.

To get an idea how the form factors are changed, we evaluate
eq.~\eqref{eq:Mdefr} with a single top partner, a single gluon
partner, and with all unknown constants fixed using a normalisation
respecting eq.~\eqref{eq:mt}.  This results in form factors behaving
as in figure \ref{fig:Mplot}.  The corrected form factor $M(p_E)$
(that provides the top quark Yukawa) is seen to decrease relative to
the uncorrected form factor at low momentum, but it then crosses zero
and continues to decrease to a magnitude bigger than the uncorrected
function.  This does not imply that the one-loop result has become
unreliable, as the expansion parameter when renormalising the
propagator goes like $\Delta m_Q/(\slashed{p}_E+m_Q)$, which is still
much less than one at large momentum.  Meanwhile the corrected form
factor $\Pi(p_E)$ displays the opposite behaviour; it is seen to
increase relative to the uncorrected form factor at low momentum, but
it decays more quickly so ends up smaller than the uncorrected
function at high momentum.

The way that these form factors appear in the Higgs potential
\eqref{eq:Vh} is model dependent so we cannot make a universal
statement about how the gluon partner correction changes the Higgs
mass.  Even for a specific model it is not obvious what will happen.
Only the magnitude of $M(p_E)$ appears in eq.~\eqref{eq:Vh}, so there
is always some cancellation when performing the momentum integral and
it is not immediately apparent whether the Higgs mass will be
increased or decreased.  A similar cancellation occurs when
integrating $\Pi(p_E)$, and the situation is further complicated by
including more top parters, whereupon there can be
cancellations in the expressions \eqref{eq:Mdefr} even before
integrating over momentum.  Nonetheless, eqs.~\eqref{eq:Deltam} and
\eqref{eq:Mdefr} are all that is needed to calculate the leading order
gluon partner correction to the Higgs mass in any model of this form.

As a specific example we now consider the minimal composite Higgs
model (MCHM)\@.  This is based on the symmetry breaking pattern
$SO(5)\to SO(4)$ and supports numerous embeddings for the top quarks.
If the left handed doublet and right handed singlet are both embedded
into ${\bf5}$'s (the MCHM$_{\bf5}$) the Higgs mass is given by
\cite{Marzocca:2012zn, Pomarol:2012qf} \be
m_h^2\approx\frac{8N_cv^2}{f^4}\int\frac{\mathrm{d}^4p_E}{(2\pi)^4}\,\sbrack{\frac{\norm{M(p_E^2)}^2}{p_E^2\Pi_L^0(p_E^2)\Pi_R^0(p_E^2)}+\pfrac{\Pi_L^h(p_E^2)}{2\Pi^0_L(p_E^2)}^2+\pfrac{\Pi_R^h(p_E^2)}{\Pi^0_R(p_E^2)}^2},
\ee where $N_c=3$ and $v=246$ GeV is the Higgs VEV.\footnote
{Here we consider the contribution to the Higgs mass from the top partners only. The one-loop contribution from the 
electroweak gauge sector, which is independent of 
the two-loop gluon correction, was estimated for this model in ref.~\cite{Pomarol:2012qf}. It is expected to be order $5\%$ of the top partner contribution.}
This expression
simplifies when the mixing between elementary and composite degrees of
freedom is small, such that $\Pi_{L/R}^0\approx1$ (equivalently
$|F|\ll m_Q$), to become \be\label{eq:mhint}
m_h^2\approx\frac{8N_cv^2}{f^4}\int\frac{\mathrm{d}^4p_E}{(2\pi)^4}\,\sbrack{\frac{1}{p_E^2}\norm{M(p_E^2)}^2+\frac{1}{4}\Pi_L^h(p_E^2)^2+\Pi_R^h(p_E^2)^2}.
\ee 
Using Weinberg sum rules \cite{Weinberg:1967kj} refs.~\cite{Marzocca:2012zn, Pomarol:2012qf} show that at least two top partners are required for the Higgs potential to be convergent in this
model.  For concreteness we will consider only two low mass states.\footnote{Including an extra layer of top partners changes the Higgs potential at the one-loop level, resulting in a different Higgs mass.  If these top partners are also light the difference can be larger than the gluon partner correction calculated here.  However, the two-point functions of the extra top partners should also be modified as in eq.~\eqref{eq:Mdefr}, so the relative importance of the overall gluon partner correction to the Higgs mass can easily remain the same.}
Denoting their masses by $m_{Q_1}$ and $m_{Q_4}$
(corresponding to their $SO(4)$ representations) the uncorrected
functions in eq.~\eqref{eq:Mdef} then take the specific forms
\begin{align}
\Pi^{h(0)}_{L/R}(p^2) &
=|F^{L/R}|^2\frac{m_{Q_4}^2-m_{Q_1}^2}{(p^2-m_{Q_4}^2)(p^2-m_{Q_1}^2)},
\nonumber\\ M^{(0)}(p^2) &
=|F^LF^R|\frac{m_{Q_4}m_{Q_1}(m_{Q_4}-m_{Q_1}e^{i\theta})}{(p^2-m_{Q_4}^2)(p^2-m_{Q_1}^2)}\sbrack{1-\frac{p^2}{m_{Q_4}m_{Q_1}}\frac{m_{Q_1}-m_{Q_4}e^{i\theta}}{m_{Q_4}-m_{Q_1}e^{i\theta}}},
\end{align}
where $\theta$ is the phase difference between form factors,
i.e.\ $F^LF^R{}^*=e^{i\theta}|F^LF^R|$ and the subscripts have been
omitted from the $F^{L/R}$ (which have been set to be equal by a field
redefinition).

Substituting into eq.~\eqref{eq:mhint} and integrating, one arrives at
the uncorrected expression \be\label{eq:mh20}
[m_h^2]^{(0)}\approx\frac{N_c}{\pi^2}\frac{m_t^2}{f^2}\frac{m_{Q_4}^2m_{Q_1}^2}{m_{Q_4}^2-m_{Q_1}^2}\ln\pfrac{m_{Q_4}^2}{m_{Q_1}^2}.
\ee To eliminate both form factors and the phase $\theta$ from the
above expression two relationships have been used, which will continue
to be assumed throughout the rest of this paper.  First one takes
$\Delta F^2=|F^L|^2-2|F^R|^2=0$.  This is helpful in arranging for
electroweak symmetry breaking, as there is an additional, positive
contribution to the quadratic term in the potential proportional to
$\Delta F^4$.  Then one uses the expression for the top mass
\be\label{eq:mt} m_t^2\approx\frac{v^2}{2f^2}|M(0)|^2, \ee which
follows from reading off the Yukawa coupling from the effective
Lagrangian \eqref{eq:Leff} in the low energy limit $p^2\approx0$.

If the renormalised expressions \eqref{eq:Mdefr} are substituted into
eq.~\eqref{eq:mhint} there is no simple analytic expression for the
functions $\Pi$ and $M$.  Nonetheless, we can still apply the two
relationships used above to solve for the $F$'s, upon which the
integrals can all be performed and the effect of the gluon partner
quantified.

\begin{figure}[!t]
\begin{center}
\includegraphics[width=0.48\textwidth]{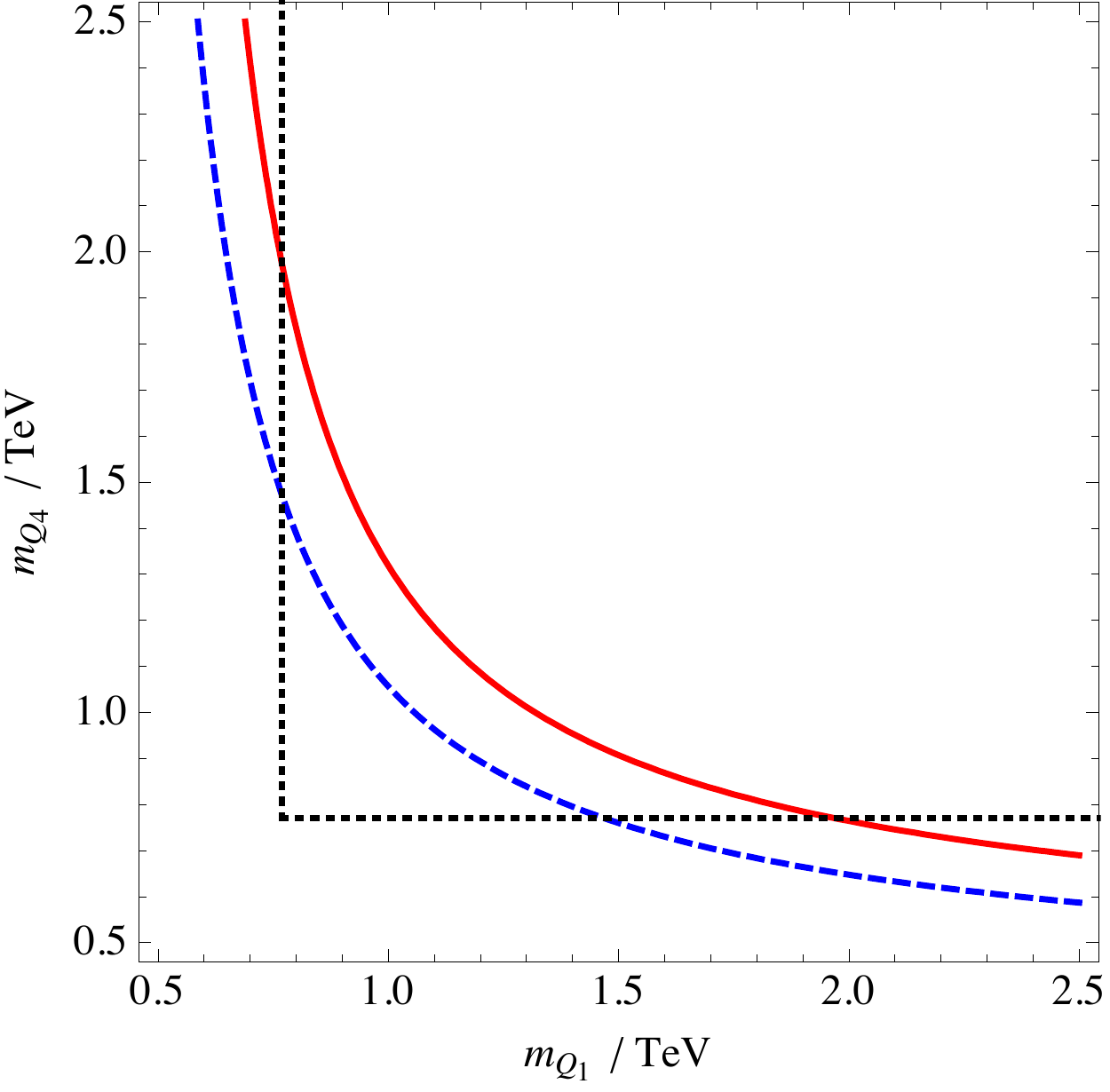}
\includegraphics[width=0.48\textwidth]{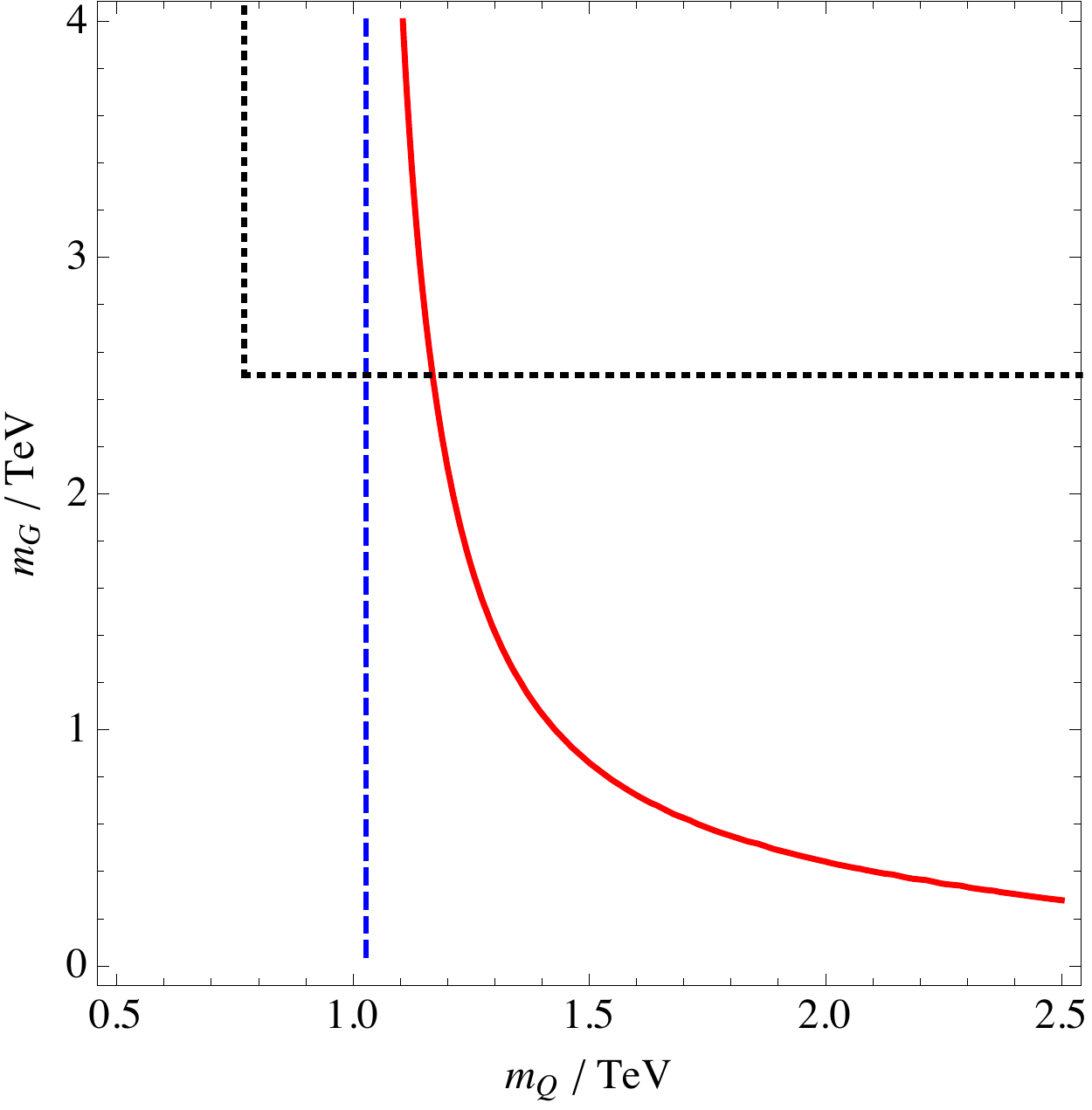}
\caption{Contours of $m_h=126$ GeV for $\alpha_G=3$, $v^2=0.1f^2$ and
  $m_t=173$ GeV\@.  Blue, dashed lines represent the result without
  massive gluons from ref.~\cite{Pomarol:2012qf} and solid, red lines
  the result including gluon partners calculated here.  Black dotted
  lines represent approximate experimental limits for the top partners
  and gluon partners of 770 GeV \cite{CMS-PAS-B2G-12-012} and 2.5 TeV
  \cite{CMS-PAS-B2G-12-006} respectively.  {\em Left}: Two different
  mass top partners with a 3 TeV gluon partner.  {\em Right}: Two
  equal mass top partners with variable gluon partner
  mass.\label{fig:Contour}}
\end{center}
\end{figure}

In figure \ref{fig:Contour} we show contours of $m_h=126$ GeV in the
$(m_{Q_1},m_{Q_4})$ plane for two distinct top partner masses.  Both
the uncorrected result from ref.~\cite{Pomarol:2012qf} (blue, dashed)
and our result, that includes the gluon partner correction (red,
solid), is shown.  For distinct top partner masses the phase $\theta$
can no longer be completely absorbed because the top mass is evaluated
at a fixed momentum but the gluon partner correction, which changes
the phase dependence, varies with momentum.  However, we have checked
that the $\theta$ dependence is only mild, so we fix $\cos{\theta}=-1$
for definiteness.  In addition, we include approximate experimental
limits for the top partners (770 GeV \cite{CMS-PAS-B2G-12-012}) and
gluon partners (2.5 TeV \cite{CMS-PAS-B2G-12-006}).

The main message of figure \ref{fig:Contour} is that the top partner
masses that result in $m_h=126$ GeV are universally shifted to larger
values.  The shift is significant, as predicted by the discussion in
section \ref{sec:NDA}, and provides additional breathing space for the
model with respect to collider searches.

\begin{figure}[!t]
\begin{center}
\includegraphics[width=0.48\textwidth]{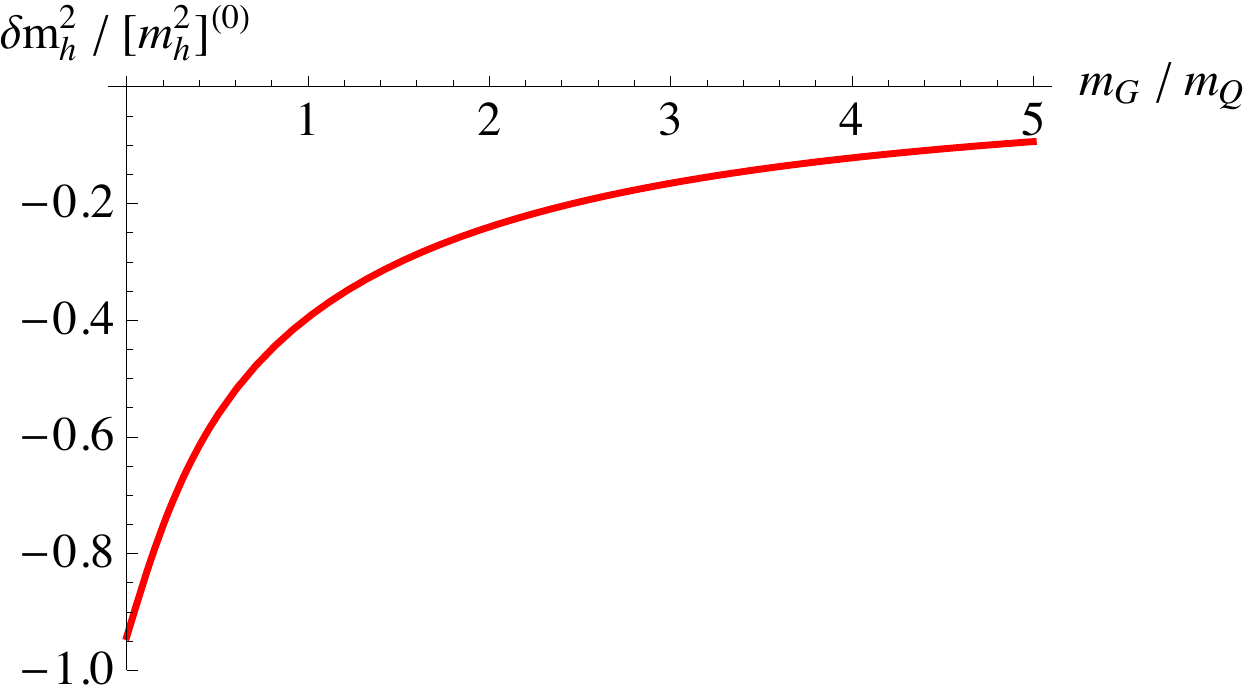}
\includegraphics[width=0.48\textwidth]{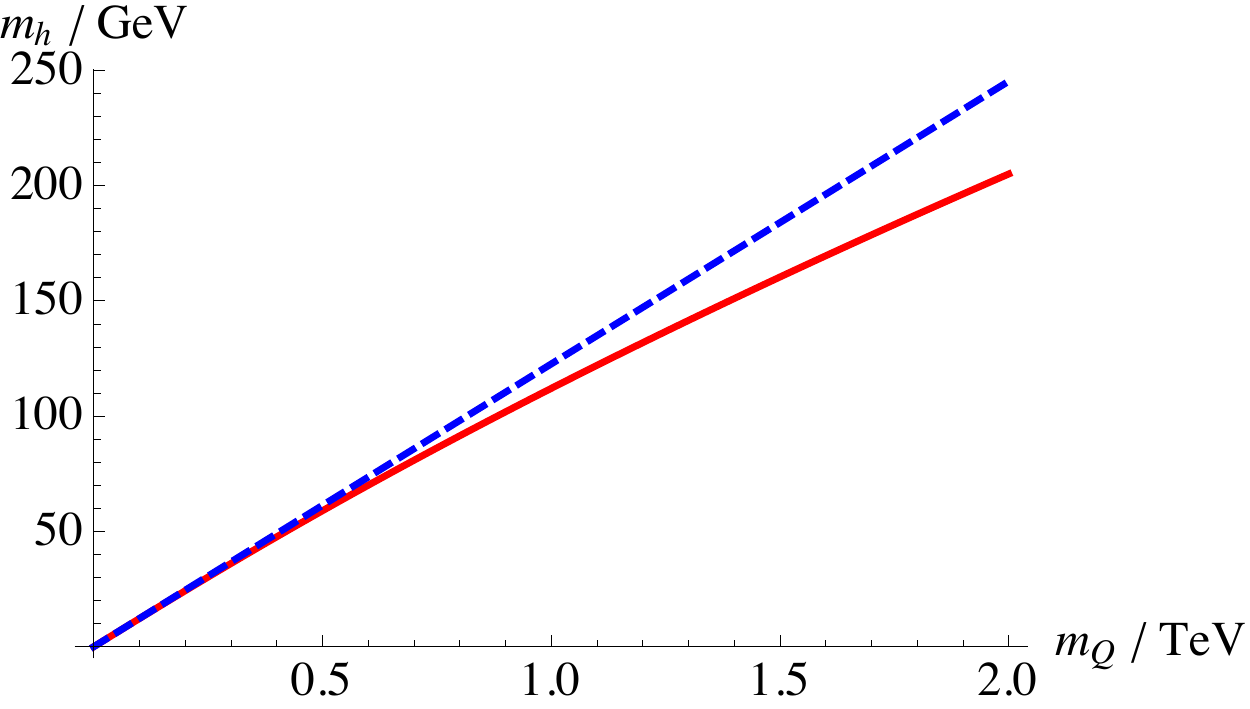}
\caption{{\em Left}: The ratio of the correction $\delta
  m_h^2=([m_h^2]^{(1)}-[m_h^2]^{(0)})$ to the uncorrected Higgs mass
  squared as a function of $m_G/m_Q$ for two equal mass top partners.
  {\em Right}: The absolute Higgs mass as a function of $m_Q$ with
  $m_G=3$ TeV in the same scenario.  The dashed line represents the
  result without massive gluons from ref.~\cite{Pomarol:2012qf} and
  the solid line the result including gluon partners calculated here.
  In both plots $\alpha_G=3$, $v^2=0.1f^2$ and $m_t=173$
  GeV.\label{fig:DeltamH2}}
\end{center}
\end{figure}

It is also instructive to consider the special case in which the two
resonances are degenerate in mass: $m_{Q_4}=m_{Q_1}\equiv m_Q$.  This
gives an uncorrected Higgs mass value of
$[m_h^2]^{(0)}=N_cm_t^2m_Q^2/\pi^2f^2$.  In figure \ref{fig:Contour}
and \ref{fig:DeltamH2} we show the effect of the gluon partner
correction in this scenario.  On the left of figure \ref{fig:DeltamH2}
is the size of the gluon partner correction relative to the
uncorrected result.  The correction is always negative (i.e.\ the
Higgs mass is decreased) and decreases in magnitude as the ratio
$m_G/m_Q$ is increased and the gluon partner decouples.  On the right
of figure \ref{fig:DeltamH2} is the absolute Higgs mass as a function
of $m_Q$ for $m_G=3$ TeV, explicitly showing the reduction of the
Higgs mass relative to the uncorrected result.  The change in the
value of the top partner mass required for a 126 GeV Higgs can be
found more precisely in this case; it is increased from 1 TeV to 1.1
TeV -- a relative change of 10\%.

\section{Conclusion}

Gluon partners are generically present in composite Higgs models that
include top partners in their low energy spectrum.  The gluon partners
are expected to couple strongly to the top partners, and this
expectation is quantitatively confirmed by arguments based on
holography.  They can therefore provide a significant correction to
the composite Higgs mass.  The leading order correction is
parameterised by a momentum dependent top partner mass shift in the
two-point functions of the strong sector, which we have explicitly
calculated in this paper.  The final effect on the Higgs mass is model
dependent, but we find a decrease in the Higgs mass in the
MCHM$_{\bf5}$.  This means that the mass of the top partners required
to yield a 126 GeV Higgs is increased by about 10\% (for a 3 TeV gluon
partner and a spontaneous global symmetry breaking scale of about 750
GeV) easing constraints from direct collider searches for top
partners.

\section*{Acknowledgements}

We thank Christophe Grojean and Alex Pomarol for useful discussions.
This work was supported in part by the Australian Research Council.
TG thanks the SITP at Stanford for support and hospitality.  JB and TG
thank the Galileo Galilei Institute for Theoretical Physics for
hospitality and the INFN for partial support during the completion of
this work.  AM thanks the KITP at Santa Barbara for hospitality.  TSR
thanks CERN Theory Division for hospitality during the final stage of
this work.  This research was supported in part by the National
Science Foundation under Grant No. NSF PHY05-51164

\appendix

\section{Holographic basis estimate}

We can also use the mixing between the holographic and mass bases to
estimate the size of the one-loop correction resulting from massive
gluons.  At next to leading order in $1/N$ there are gluon partner
contributions to the two-point functions of $\cO_L$ and $\cO_R$.  For
simplicity, let us assume that the strong sector only produces one
gluon partner, $G_c$, that mixes with the elementary gluon, $A_e$, so
that the mass eigenstates can be written as
\begin{align}
A_g^\mu & =A^\mu_e\cos{\theta}+G_c^\mu\sin{\theta}, & G^\mu &
=G_c^\mu\cos{\theta}-A^\mu_e\sin{\theta},
\end{align}
where $A_g$ is the (massless) physical gluon and $G$ is the massive
gluon.  There is then one diagram for each mass eigenstate propagating
around the loop.  Each diagram gets a factor of $\alpha_G$ from the
$\bar{Q}\slashed{G}_cQ$ coupling in the strong sector, and a
$\sin^2\theta$ or $\cos^2\theta$ from the $G_c$-$A_e$ mixing to give
\begin{align}
\Sigma_g & \sim\frac{N\alpha_G}{(4\pi)^3}\sin^2\theta\, C_2(N_c), &
\Sigma_G & \sim\frac{N\alpha_G}{(4\pi)^3}\cos^2\theta\, C_2(N_c),
\end{align}
where $C_2(3)=4/3$ is the quadratic Casimir for the fundamental
representation of $SU(3)$ and $\Sigma_{g,G}$ is multiplied by an order
one loop function.  The factor of $N$ comes from factors of
$\sqrt{N}/(4\pi)$ on each external leg, themselves originating from
vacuum creation amplitudes.  Although $\alpha_G$ scales like $1/N$ it
can still be large if this scaling comes with a large prefactor.

In the case at hand we know from gauge invariance that
$\alpha_G\sin^2\theta=\alpha_3$.  We also know from holographic
arguments that $\sin^2\theta\approx1/\pi kR\approx1/30$ (see
e.g.\ ref.~\cite{Gherghetta:2010cj} or ref.~\cite{Carena:2007tn} for
an explicit calculation).  Hence $\alpha_G\approx30\alpha_3$, clearly
overcoming any $1/N$ suppression, and we find \be
\Sigma_G\sim\frac{10N\alpha_3}{16\pi^3}.  \ee The ratio of the
one-loop result with the tree-level result $\sim N/(16\pi^2)$ gives a
factor $1/\pi$ , which is the same as obtained in section
\ref{sec:NDA}.

\section{Passarino-Veltman integrals}

Expressions for the Passarino-Veltman integrals used are
\begin{align}
B_0(p^2,m_Q^2,m_G^2) &
=\frac{1}{i\pi^2}\dimint{4}{k}\frac{1}{[k^2-m_Q^2][(k-p)^2-m_G^2]}
\nonumber\\ &
=\Delta_\epsilon-\dimintlim{}{x}{0}{1}\ln\sbrack{\frac{xm_G^2+(1-x)m_Q^2-x(1-x)p^2}{\mu^2}},
\\ B^\mu(p^2,m_Q^2,m_G^2) &
=\frac{1}{i\pi^2}\dimint{4}{k}\frac{k^\mu}{[k^2-m_Q^2][(k-p)^2-m_G^2]}
\nonumber\\ & = p^\mu B_1(p^2,m_Q^2,m_G^2) \nonumber\\ &
=p^\mu\nbrack{-\frac{1}{2}\Delta_\epsilon+\dimintlim{}{x}{0}{1}x\ln\sbrack{\frac{xm_G^2+(1-x)m_Q^2-x(1-x)p^2}{\mu^2}}},
\end{align}
where $\mu$ is the renormalisation scale.

\providecommand{\href}[2]{#2}\begingroup\raggedright\endgroup

\end{document}